\documentclass{PoS}
\usepackage[]{gensymb}
\usepackage{enumerate}
\newcommand{\eh}[1]{\,\mathrm{#1}}

\newcommand{\tev}{\eh{TeV}}

\newcommand{\kyr}{\eh{kyr}}

\newcommand{\pc}{\eh{pc}}

\newcommand{\ergs}{\eh{erg\,s^{-1}}}

\newcommand{\tin}[1]{_\mathrm{#1}}

\newcommand{\dg}{^{\circ}}

\newcommand{\che}[1]{}

\renewcommand{\epsilon}{\varepsilon}

\newcommand{\hess}{H.E.S.S.}

\newcommand{\hgps}{H.E.S.S. Galactic Plane Survey}

\newcommand{\scatter}{\emph{varied model}}

\newcommand{\hessetal}{H.E.S.S. Collaboration {\em et al.}}
\newcommand{\pwnpop}{H.E.S.S. Collaboration {\em et al.} 2017 \cite{pwnpop}}

\let\seriesfb\bfseries\def\bfseries{\boldmath\seriesfb}
\let\seriesdm\mdseries\def\mdseries{\unboldmath\seriesdm}

\newcommand{\velax}{{Vela\,X}}

\newcommand{\lmc}{{N\,157B}}

\newcommand{\threec}{3C\,58}

\newcommand{\edot}{\dot{E}}

\newcommand{\age}{\tau\tin{c}}
\newcommand{\lumi}{L_{1-10\,\mathrm{TeV}}}

\newcommand{\fref}[1]{Fig.~\ref{#1}}

\title{The population point of view on the evolution of \\  TeV pulsar wind nebulae}

\ShortTitle{Evolution of TeV pulsar wind nebulae}

\author{\speaker{S.~Klepser} $^a$,
Y.~Gallant$^b$,
M.~Mayer$^c$,
K.~Valerius$^d$
for the \hess\ collaboration\\
\llap{$^a$}DESY, D-15738 Zeuthen, Germany \\
\llap{$^b$}Laboratoire Univers et Particules de Montpellier, Universit\'e Montpellier, CNRS/IN2P3,  CC 72, Place Eug\`ene Bataillon, F-34095 Montpellier Cedex 5, France \\
\llap{$^c$}Institut f\"ur Physik, Humboldt-Universit\"at zu Berlin, Newtonstr. 15, D 12489 Berlin, Germany \\
\llap{$^d$}Karlsruhe Institute of Technology, Institute for Nuclear Physics, P.O. Box 3640, D-76021 Karlsruhe, Germany \\
}

\abstract{
To investigate the nature and evolution of TeV pulsar wind nebulae, we examine the firmly identified PWNe in the H.E.S.S. Galactic Plane Survey, along with the few other known detections from the literature, as well as the upper limits extracted from the H.E.S.S survey. These data exhibit a correlation of TeV surface brightness with pulsar spin-down power. It appears to be caused by both an increase of TeV extension and a decrease of TeV luminosity with decreasing spin-down power. We also find that the offsets of pulsars with ages around 10 kyr with respect to the wind nebula centres are frequently larger than can be plausibly explained by pulsar proper motion and could be due to an asymmetric environment. These and other results will be presented and put to context with a basic modelling of TeV pulsar wind nebula evolution.
}

\FullConference{35th International Cosmic Ray Conference --- ICRC2017\\
		10--20 July, 2017\\
		Bexco, Busan, Korea}

\begin{document}
This conference contribution is based on the recent paper on the
subject
(\pwnpop). All details, more plots, and references can be found in that
reference. This proceedings paper wraps up the main conclusions and key figures of the
paper. 

\section{Overview}

The paper subsumes and examines the population of TeV pulsar wind nebulae
(PWNe) found to date. An updated census presents 14 objects reanalysed in the \hgps\
(HGPS) pipeline, which are considered to be firmly identified PWNe. Five more
objects could be found
outside that catalogue range or pipeline. In an evaluation of candidate PWNe,
we conclude
that there
are ten strong further candidates in the HGPS data.

\begin{figure}
\centerline{\includegraphics[width=0.7\linewidth]{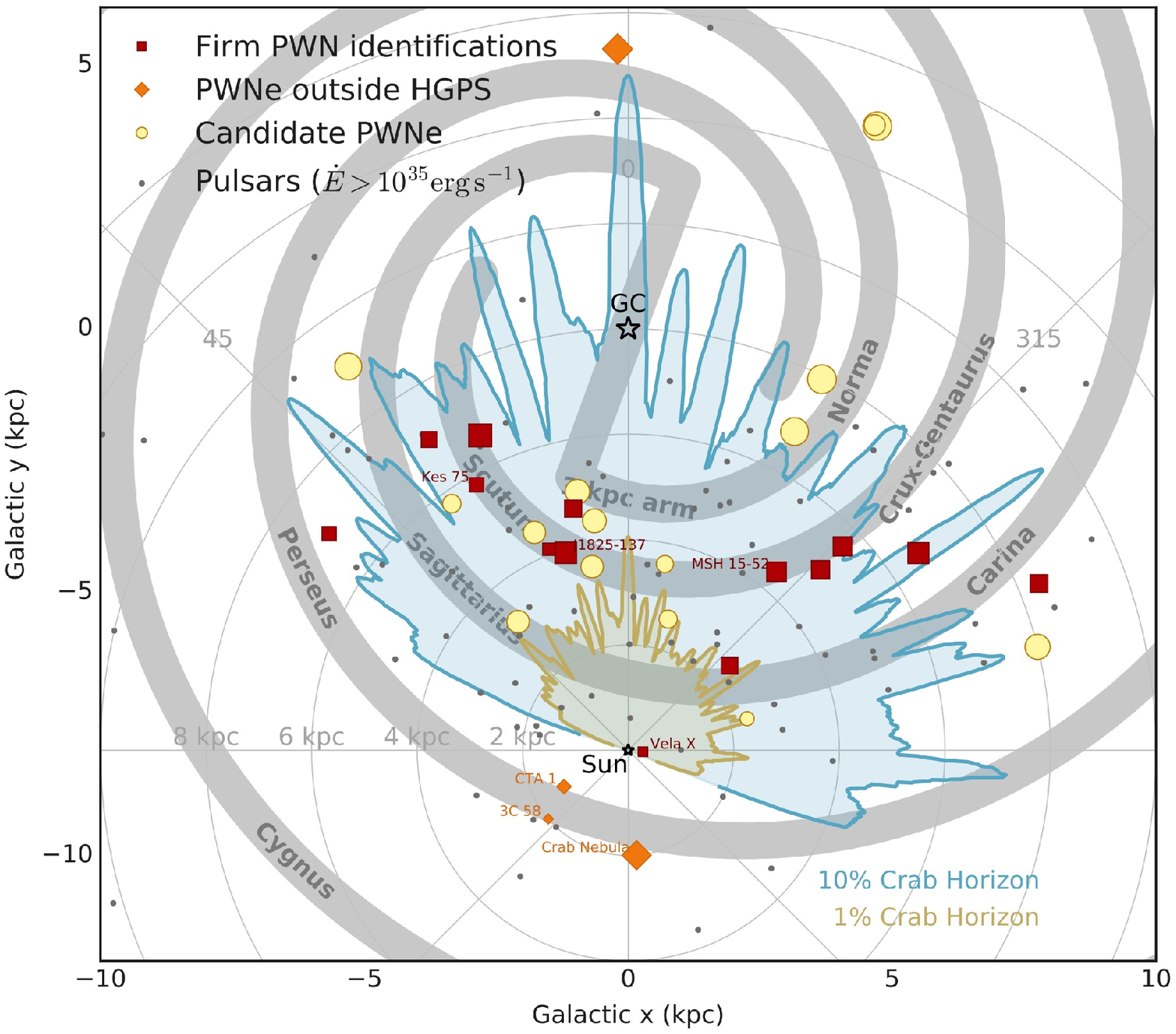}}
\centerline{\includegraphics[width=0.7\linewidth]{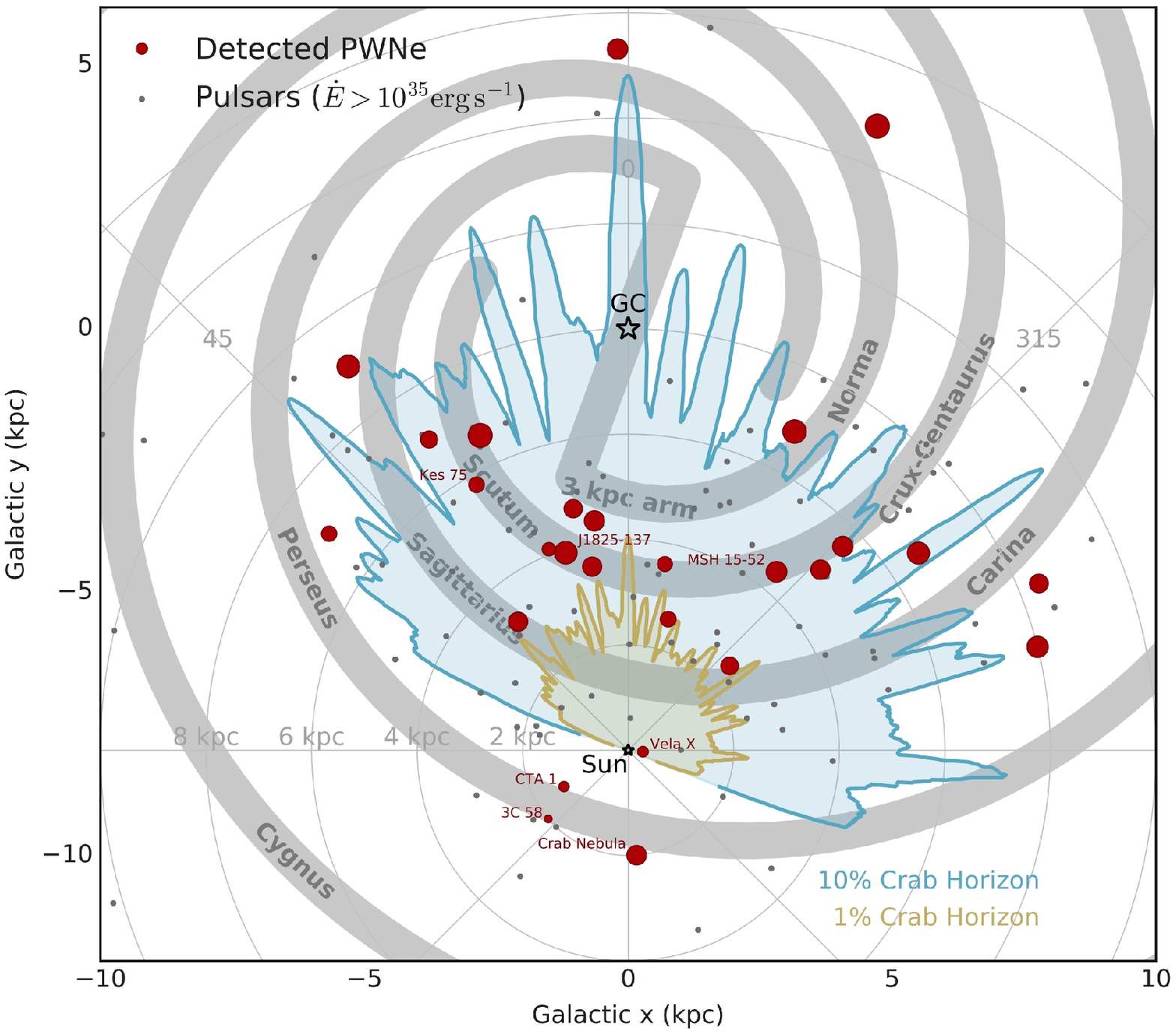}}
\caption{Schematic of the objects discussed here in the context of the Milky Way and its spiral arms. The yellow and blue curves outline the sensitivity horizon of the HGPS for 
point-like sources with an integrated gamma-ray luminosity ($1$--$10\tev$) 
of 1\% and 10\% of the Crab luminosity, respectively. Top: Firmly identified
PWNe, candidates, and energetic pulsars ($\edot > 10^{35}\ergs$) without
detected TeV wind nebula. Bottom: Simplified view with all firm (HGPS and external) identifications and  positively rated candidates displayed with the same symbol, and negative candidates removed.
The figures are reproduced from \hessetal\ 2017$^1$, but the sizes of the symbols are scaled with $\log{\lumi}$} here.
\label{fig:pwn_galaxy}
\end{figure}

Most of the PWNe are located in the bright and dense Crux Scutum arm of the
inner Milky Way (\fref{fig:pwn_galaxy}). 
A spatial correlation study 
confirmed the picture
drawn in earlier studies, namely that only
young, energetic pulsars grow TeV pulsar wind nebulae that are bright enough
for
detection with presently available Cherenkov telescopes.
For the first time, flux upper limits for undetected PWNe are given
around 22 pulsars with a spin-down power beyond
$10^{35}\ergs$ and with expected apparent extensions (plus
offsets) below $0.6\dg$ in the sky.

\section{Census of PWNe around high-$\edot$ pulsars}

Of the 17 most energetic ATNF pulsars, with a spin-down power
of $\edot \geq 10^{37}\ergs$, 11 have either an identified TeV
wind nebula (9) or candidate (2) featured in
the present study. Of the remaining 6,
\begin{itemize}
        \item 3 are included in Table~5 in \pwnpop, where all flux limits of
pulsars without detected PWN are listed;
        \item 3 are out of the range of the HGPS:
        \begin{itemize}
           \item PSR J2022+3842: SNR~G076.9+01.0, contains an X-ray PWN; not
reported in TeV
           \item PSR J2229+6114: Boomerang, contains an X-ray PWN; detected by
MILAGRO and VERITAS, but of
unclear nature in TeV
           \item J0540$-$6919: In the Large Magellanic Cloud; a limit is given
in \hessetal\ 2015 \cite{hess_lmc_science}. 
        \end{itemize} 
\end{itemize} 
Concluding, only 5 of the 17 highest-$\edot$ pulsars remain without a detected
potential counterpart in the TeV band. 

\section{PWN evolution}

\begin{figure}
\begin{minipage}{0.48\linewidth}
\centerline{\includegraphics[width=1.0\linewidth]{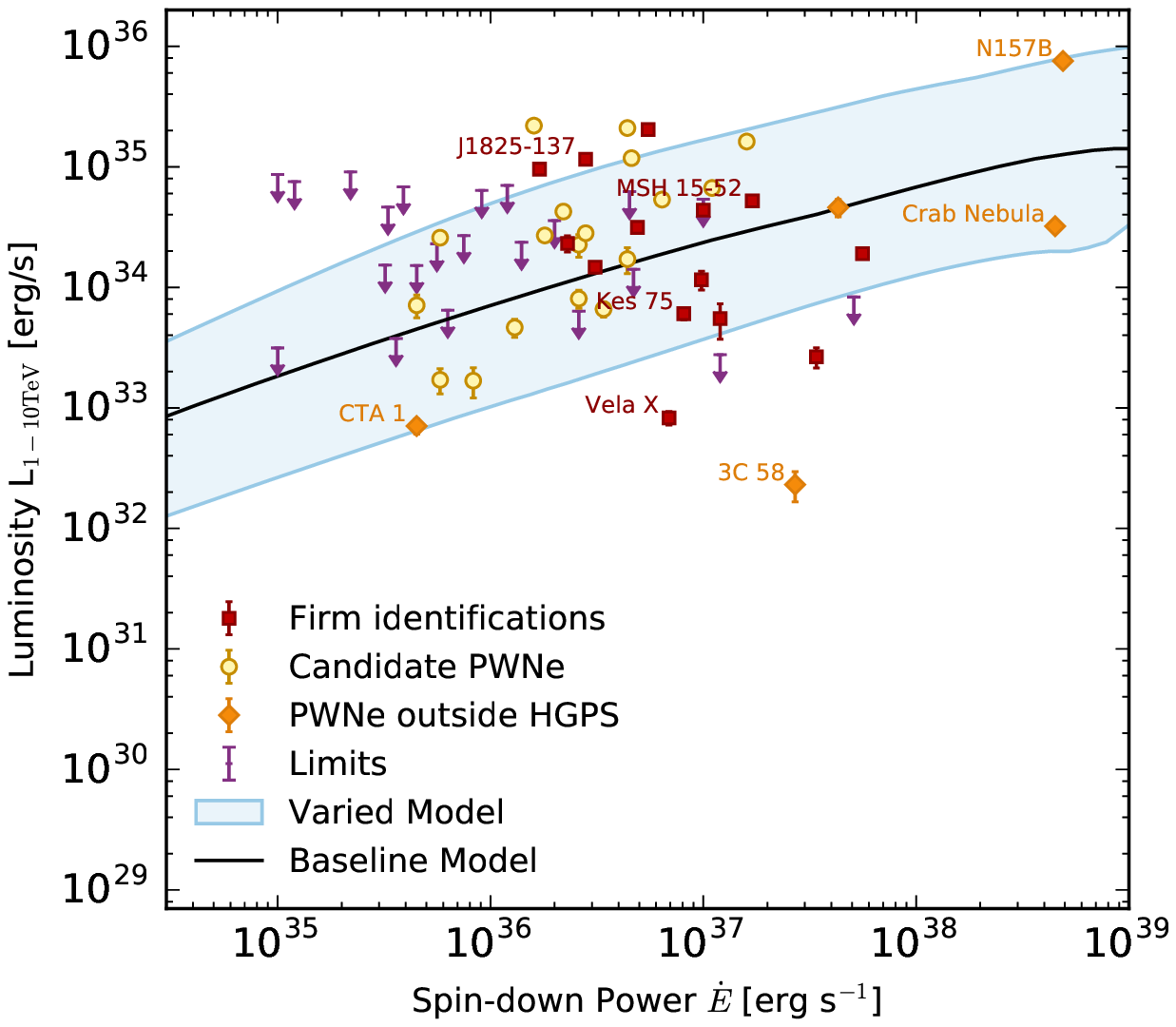}}
\end{minipage}
\hfill
\begin{minipage}{0.48\linewidth}
\centerline{\includegraphics[width=1.0\linewidth]{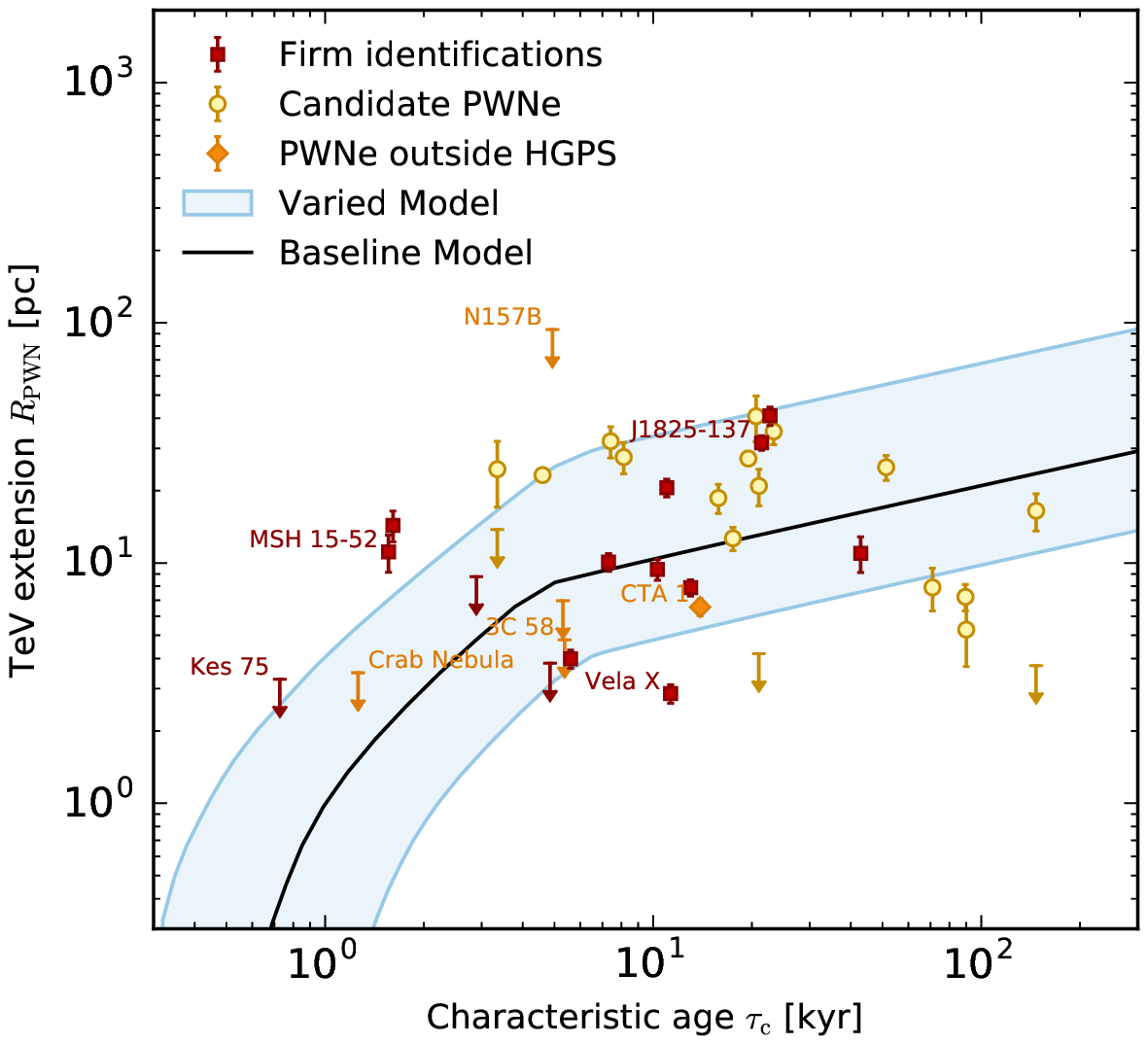}}
\end{minipage}
\caption[]{Left: Relation of TeV luminosity and
pulsar $\edot$. Right: PWN extension evolution with time, in comparison to the modelling 
considered in this work. The figures are reproduced from
\pwnpop.}
\label{fig:lum_ext}
\end{figure}

\begin{figure}
\begin{minipage}{0.48\linewidth}
\centerline{\includegraphics[width=1.0\linewidth]{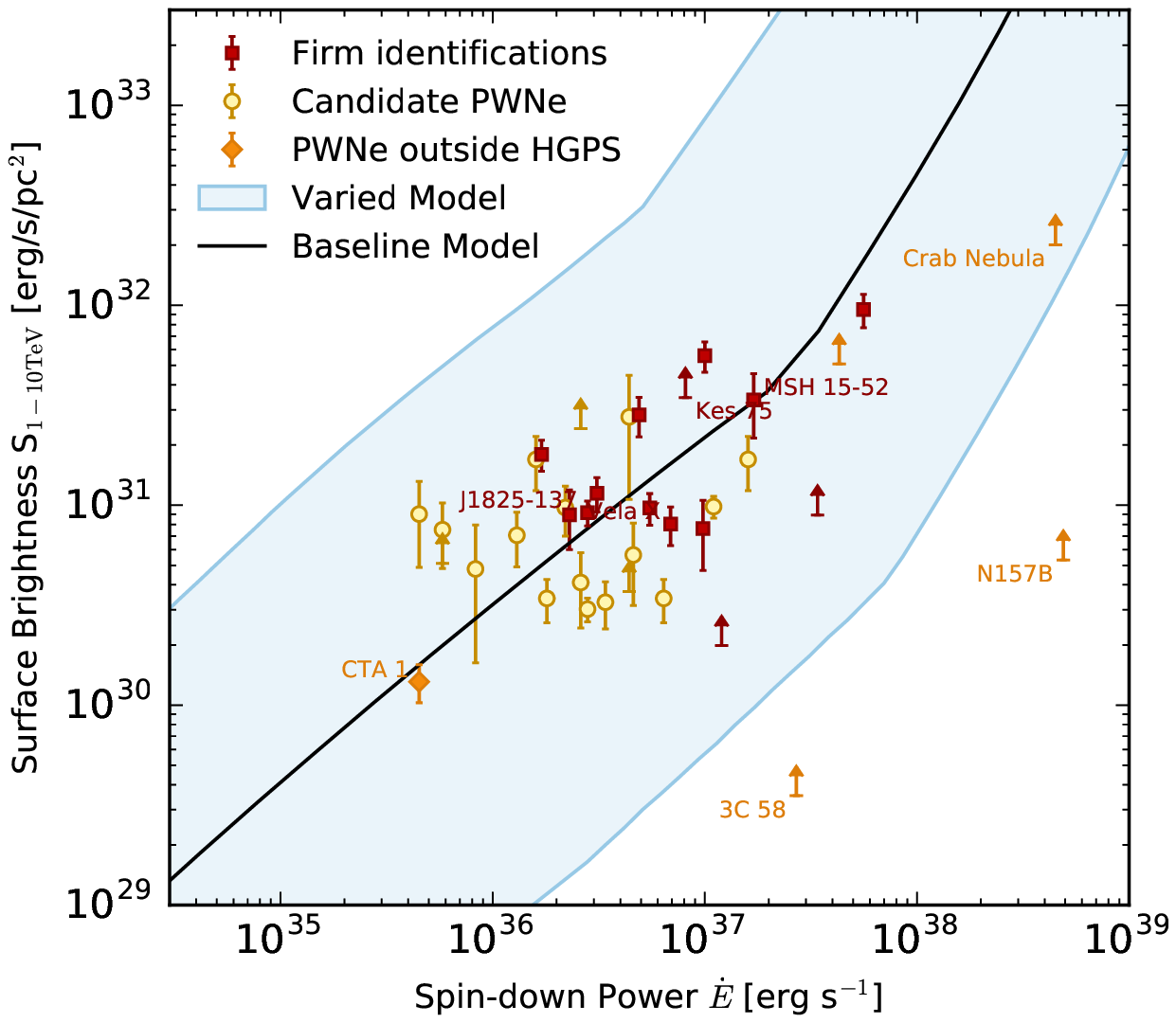}}
\end{minipage}
\hfill
\begin{minipage}{0.48\linewidth}
\centerline{\includegraphics[width=1.0\linewidth]{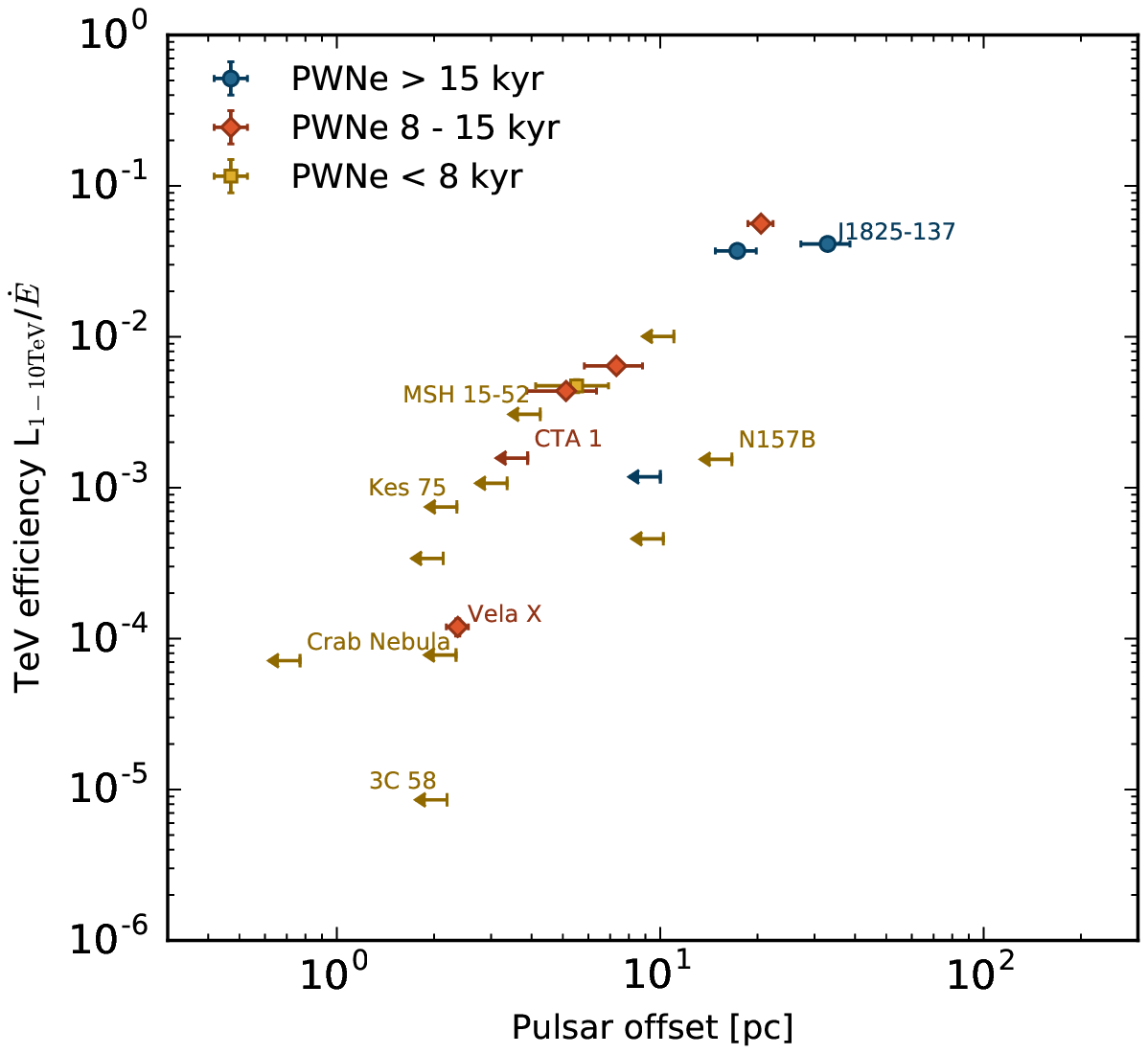}}
\end{minipage}
\caption[]{Left: Relation of TeV surface brightness and
pulsar $\edot$. Right: TeV efficiency as a function of pulsar
offset, plotted for pulsars of different age groups. High-offset systems tend
to be
more TeV-efficient than low-offset systems. The figures are reproduced from
\pwnpop.}
\label{fig:sb_off}
\end{figure}

Figures~5 to 10 in \pwnpop\ show, like \fref{fig:lum_ext} and \fref{fig:sb_off}
in this proceedings paper, a variety of
trends between pulsar and TeV wind nebula parameters, and consistently
compare them to a simple one-zone time-dependent emission model of the TeV
emission with a varied range of model input parameters. The
main conclusion is that for several observables, a trend was
found in the data that is consistent with the trends suggested by our
model. With a moderate variation of
the model input parameters, we can mimic also the spreads of the observables.
Our first-order understanding of the evolution of TeV pulsar wind
nebulae with ages up to some tens of kiloyears therefore seems to be
compatible with what the whole population of detected and
undetected PWNe suggests.

More concretely, using the flux limits for undetected PWNe, we find evidence that the TeV
luminosity
of PWNe decays with time while they expand in
size, preventing the detection of those whose pulsar has dropped 
below a spin-down of $\sim 10^{36}\ergs$ (roughly corresponding to several tens of
kiloyears). 
This was implicitly known before from the mere non-detection of
old TeV pulsar wind nebulae, but for the first time could be put into a
quantitative perspective, both by fitting data and limits, and by
comparing the
data to model predictions. The power-law relation between TeV luminosity and
pulsar spin-down could be estimated as $\lumi\sim\edot^{0.58\pm0.21}$,
in consistency with the model, which suggests a power index of around $0.5$.

Another feature that was discussed on some individual objects before 
(e.g. \cite{hess_j1825_detection}) is the ``crushing"
of PWNe, which can be exerted by the inward-bound reverse front of the
supernova shock wave. For SNRs that develop asymmetrically, for instance due to
an inhomogeneous surrounding medium, this crushing may result in
distortion and displacement of the wind nebula. A very bright, very extended example of this is HESS~J1825$-$137, also presented at this conference \cite{alison_j1825}. Put to a population-scoped
context, it becomes clear that pulsar proper motions alone
are not sufficient to
explain the large offsets observed, and some kind of a crushing mechanism may
indeed be the dominant and frequent cause of pulsar-PWN offset in middle-aged systems.
Furthermore, the offsets of PWNe from their pulsars appear to relate to high efficiency
(\fref{fig:sb_off}, right), suggesting that the PWNe
either gain energy and brightness through the process that causes the offset
or that dense
surroundings amplify both the IC luminosity and the offset between pulsar and
wind nebula. While the evidence for this at present is not very strong,
following
up with expanded future studies is certainly worthwhile. 

The expansion of PWNe with time was also shown to be evident in the data. The fitted
relation $R\sim\age^{0.55 \pm 0.23}$ suggests an average expansion coefficient
in between those expected in theory ($1.2$ and $0.3$). The data set is not
comprehensive enough to do a fit with two power laws, but appears to be
consistent with the model (\fref{fig:lum_ext}, right). Notably, this expansion is not so clear in X-rays, where the
synchrotron
emission always remains very local because it only traces the young particles
in areas of high magnetic field
relatively close to the pulsar. Most of the old objects ($>30\kyr)$
are therefore smaller than $1\pc$ in their bright X-ray core emission.

As a consequence of the
two moderate correlations of luminosity and spatial extent with pulsar
$\edot$, a stronger correlation was found between the PWN surface brightness and 
pulsar $\edot$ (\fref{fig:sb_off}, left). What stands out is not
only the correlation itself, but also its relatively low scatter.

The evolution trend of the photon index remains an open issue at present.
Neither the data
nor the model are particularly clear about it for the young to middle-aged
PWNe we investigated.

\section{Detection biases}
Since both the \hgps\ and the ATNF pulsar database only cover a fraction of
the
Milky Way, depending on TeV and pulsar brightnesses, the study in
\pwnpop\ suffers from
several selection biases discussed throughout the paper. For TeV-bright,
high-$\edot$, young pulsar systems ($>10^{36}\ergs$) we achieve a relatively
good coverage, whereas for systems beyond some tens of kiloyears of age we
likely
miss many sources. In the plots discussing flux-related quantities, this is
partly compensated by the inclusion of flux limits, allowing for statements
that consider the presence of non-detections. For extension- and position-related quantities,
however, we can only rely on the detected cases.
It would require a full
population synthesis study to judge whether some of the correlations are
genuine or
include side effects of other correlations or selection biases. This usually
needs many astrophysical assumptions and theoretical suppositions, which was beyond the scope of this
experimental paper.

One presumably very influential parameter ignored in this study is the density
of matter and background light at the position of each pulsar. It is likely
due to such circumstances that \threec, CTA~1 and \velax, also presented at this conference \cite{luigi_velax} are so faint (see \fref{fig:pwn_galaxy}), and
\lmc\ (in the
Large Magellanic Cloud) is so
bright. In the scope of a population synthesis study, one could use a
specific Milky Way model to ``calibrate" the calorimetric objects that
TeV pulsar wind nebulae are assumed to be.

\section{Modelling}

\begin{figure}
\begin{minipage}{0.48\linewidth}
\centerline{\includegraphics[width=1.0\linewidth]{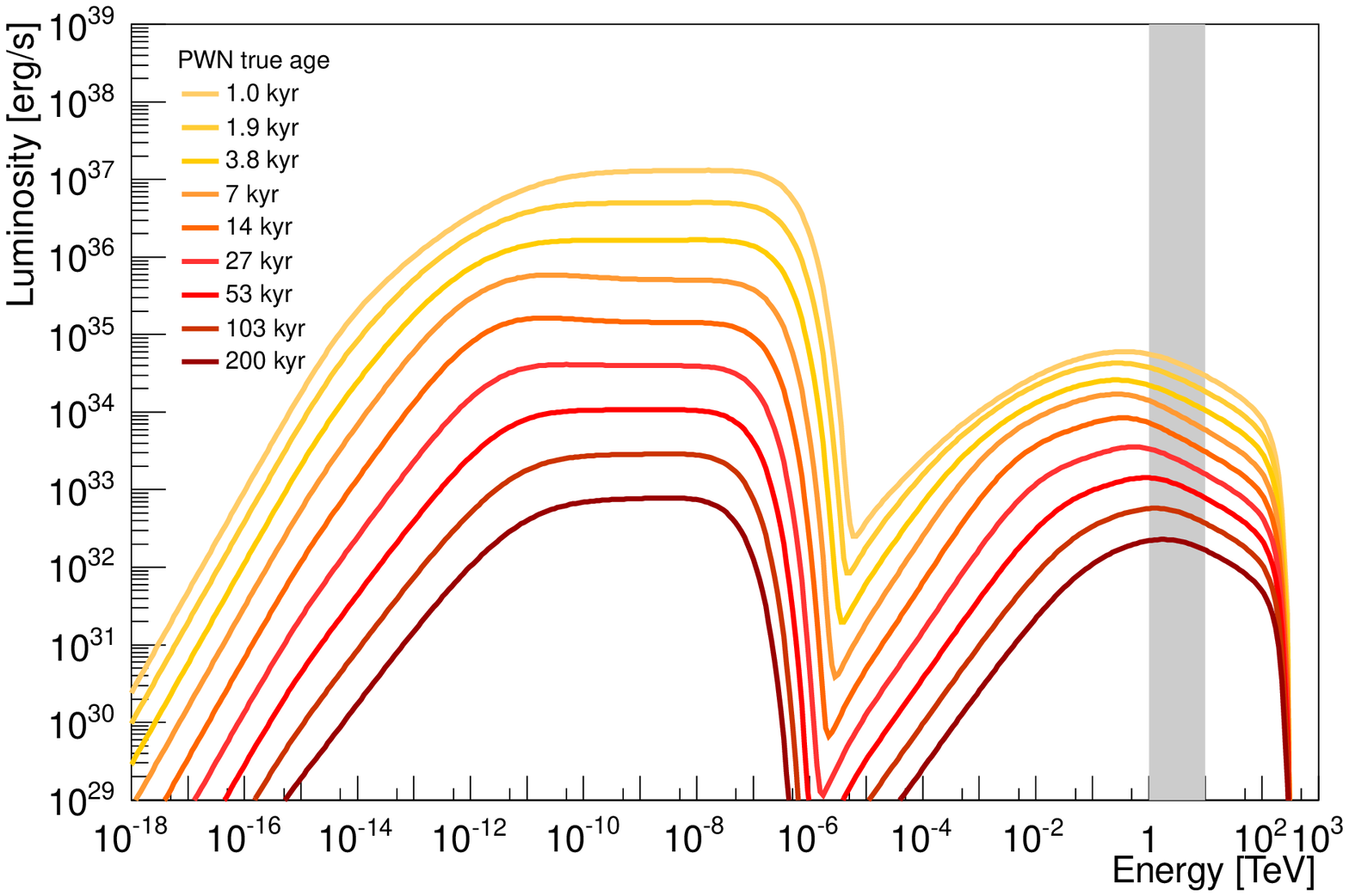}}
\end{minipage}
\hfill
\begin{minipage}{0.48\linewidth}
\centerline{\includegraphics[width=1.0\linewidth]{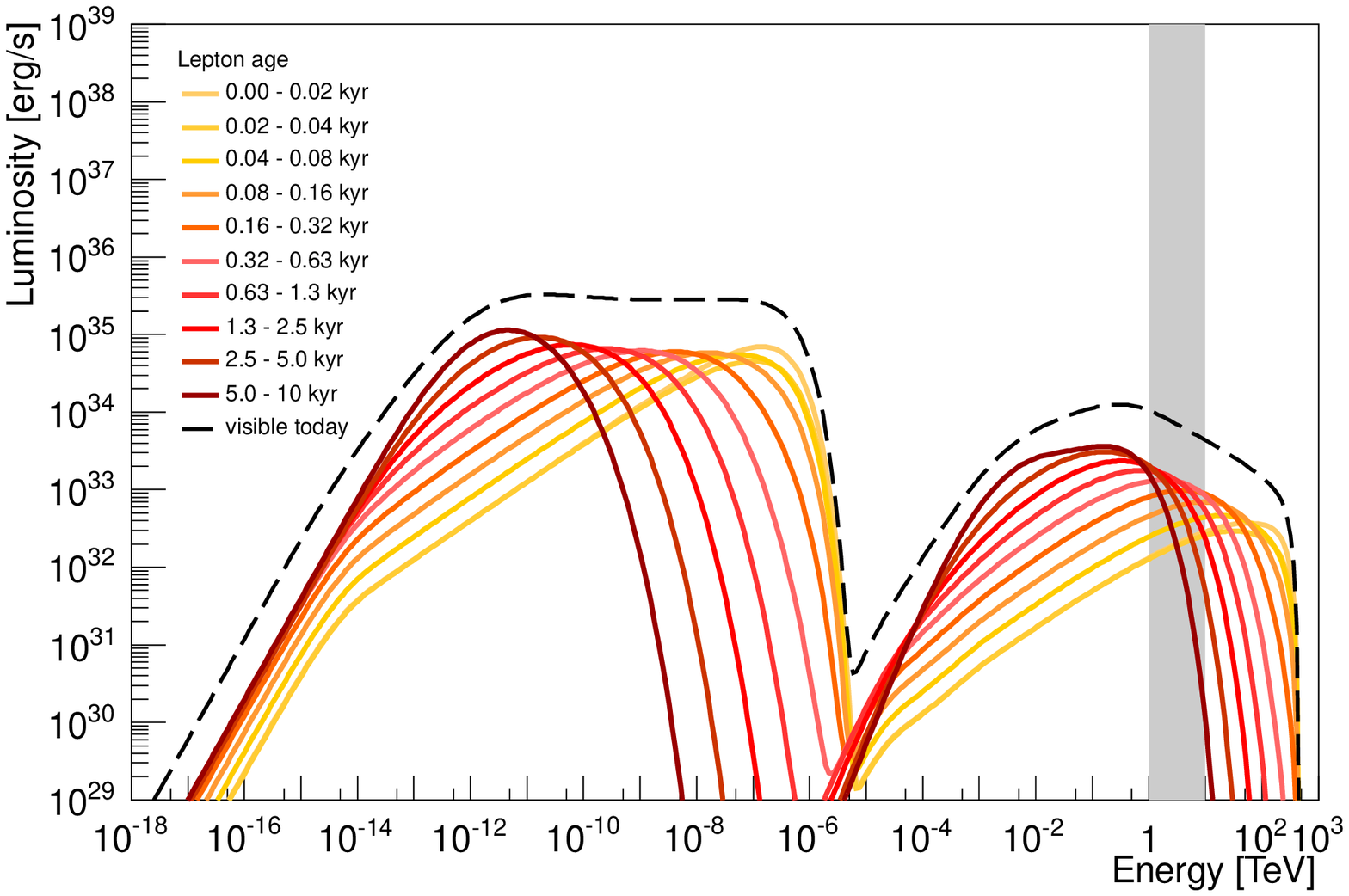}}
\end{minipage}
\caption[]{Modelled spectral energy distribution (SED) of a generic PWN
according to the model given in \cite{pwnpop}, Appendix A.
Left: Time evolution of the SED, ranging from $1\kyr$ to $200\kyr$.
Right: Decomposition of the SED of a 
middle-aged PWN ($10\kyr$; black dashed curve) into contributions
by leptons from various injection epochs (coloured lines).
The grey-shaded bands indicate the energy range of $1$--$10\tev$ explored in
this work. The figures are reproduced from \pwnpop.
}
\label{fig:model}
\end{figure}

On the modelling side, we are able to describe the trends and scatter of the
TeV properties of the present
PWN population with a relatively simple time-dependent modelling described in
Appendix A of \pwnpop\ and whose basic evolution is displayed in
\fref{fig:model}. Its $12$
free parameters
($7$ of which were varied for the \scatter) were well below the $4\times
19$ observed
parameters that the firmly identified PWNe provided.
It is remarkable that the adaptive parameters needed to be varied in a
fairly small range, compared to what one may fathom
from the modelling literature, while still producing sufficient scatter in
the predicted observables. Whether this indicates that the
variations of the individual PWN parameters are indeed small, or
whether this is an effect of the parameters being (anti-)correlated (see caveats
discussion in A.7 of \pwnpop), could not be clarified in this work. It might
require a
deeper physical model of the pulsars and possibly a multidimensional
likelihood fit to correctly quantify all correlations and
identify the true distributions of its parameters.

\section{Outlook}

In the CTA era, many of the PWNe that will be detected in addition to the
now assessed population will be middle-aged and old systems that are too
faint or too extended to be detected with current instruments.
Also, improvements in the radio and gamma-ray band coverage will enlarge
the sample of pulsars detected in our Galaxy.
To gain new insights from studying these systems, a solid and publicly
available modelling code is needed that includes the
difficult reverse shock interaction phase of a PWN in a reproducible
way. 
This may help to understand the effect and influence of the amount of
crushing and pulsar offset of the PWN, which is likely an 
influential factor of later PWN evolution.

On the analysis side, it would be beneficial to (i) improve the angular
resolution and get to smaller scales of extension, (ii) find ways to
reliably disentangle overlapping sources and their spectra, and (iii) aim for
detecting objects larger than the IACT camera FOV. It is only if this is
improved that larger datasets and more exposure can help us to
unriddle sources that are closeby or occult each other in
the densely populated arms of the Galaxy.

\section*{Acknowledgments}

\small
The support of the Namibian authorities and of the University of Namibia in
facilitating the construction and operation of H.E.S.S. is gratefully
acknowledged, as is the support by the German Ministry for Education and
Research (BMBF), the Max Planck Society, the German Research Foundation (DFG),
the French Ministry for Research, the CNRS-IN2P3 and the Astroparticle
Interdisciplinary Programme of the CNRS, the U.K. Science and Technology
Facilities Council (STFC), the IPNP of the Charles University, the Czech
Science Foundation, the Polish Ministry of Science and Higher Education, the
South African Department of Science and Technology and National Research
Foundation, the University of Namibia, the Innsbruck University, the Austrian
Science Fund (FWF), and the Austrian Federal Ministry for Science, Research
and Economy, and by the University of Adelaide and the Australian Research
Council. We appreciate the excellent work of the technical support staff in
Berlin, Durham, Hamburg, Heidelberg, Palaiseau, Paris, Saclay, and in Namibia
in the construction and operation of the equipment. This work benefitted from
services provided by the H.E.S.S. Virtual Organisation, supported by the
national resource providers of the EGI Federation.


\normalsize


\begin{thebibliography}{99}
\bibitem{pwnpop}H.E.S.S. Collaboration et al., accepted for publication in
{\em A\&A}, arXiv:1702.08280.

\bibitem{hess_lmc_science} H.E.S.S. Collaboration et al., {\em Science}~{\bf 347}~{406} (2015), arXiv:1501.06578.

\bibitem{hess_j1825_detection} A.F. Aharonian et al. (H.E.S.S. Collaboration), {\em A\&A}~{\bf 442}~{L25} (2005), arXiv:astro-ph/0510394.

\bibitem{alison_j1825} A.M.W. Mitchell et al., \pos{PoS(ICRC2017)707}

\bibitem{luigi_velax} L. Tibaldo et al., \pos{PoS(ICRC2017)719}

%
%
%

\end{thebibliography}
\end{document}